\documentclass{aa}
\usepackage{graphicx,color,amssymb}
\usepackage{natbib}
\usepackage{soul}

\begin{document}
\title{Damping of prominence longitudinal oscillations due to mass accretion}

\author{Michael S. Ruderman\inst{1,2} \and Manuel Luna\inst{3,4}}
\institute{School of Mathematics and Statistics (SoMaS), The University of Sheffield, Hicks Building, Hounsfield Road, Sheffield S3 7RH, United Kingdom \and
Space Research Institute (IKI) Russian Academy of Sciences, Moscow, Russia \and
Instituto de Astrof{\'{\i}}sica de Canarias, E-38200 La Laguna, Tenerife, Spain \and Universidad de La Laguna, Dept. Astrof{\'{\i}}sica, E-38206 La Laguna, Tenerife, Spain}

\date{Received / Accepted}

\abstract{We study the damping of longitudinal oscillations of a prominence thread caused by the mass accretion. We suggested a simple model describing this phenomenon. In this model we considered a thin curved magnetic tube filled with the plasma. The prominence thread is in the central part of the tube and it consists of dense cold plasma. The parts of the tube at the two sides of the thread are filled with hot rarefied plasma. We assume that there are flows of rarefied plasma toward the thread caused by the plasma evaporation at the magnetic tube footpoints. Our main assumption is that the hot plasma is instantaneously accommodated by the thread when it arrives at the thread, and its temperature and density become equal to those of the thread. Then we derive the system of ordinary differential equations describing the thread dynamics. 

We solve this system of ordinary differential equations in two particular cases. In the first case we assume that the magnetic tube is composed of an arc of a circle with two straight lines attached to its ends such that the whole curve is smooth. A very important property of this model is that the equations describing the thread oscillations are linear for any oscillation amplitude. We obtain the analytical solution of the governing equations. Then we obtain the analytical expressions for the  oscillation damping time and periods. We find that the damping time is inversely proportional to the accretion rate. The oscillation periods increase with time. We conclude that the oscillations can damp in a few periods if the inclination angle is sufficiently small, not larger that $10^\circ$\/, and the flow speed is sufficiently large, not less that 30~km/s. 

In the second model we consider the tube with the shape of an arc of a circle. The thread oscillates with the pendulum frequency dependent exclusively of the radius of curvature of the arc. The damping depends on the mass accretion rate and the initial mass of the threads, that is the mass of the thread at the moment when it is perturbed. First we consider small amplitude oscillations and use the linear description. Then we consider nonlinear oscillations and assume that the damping is slow, meaning that the damping time is much larger that the characteristic oscillation time. The thread oscillations are described by the solution of the nonlinear pendulum problem with slowly varying amplitude. The nonlinearity reduces the damping time, however this reduction is small. Again the damping time is inversely proportional to the accretion rate. We also obtain that the oscillation periods decrease with time. However even for the largest initial oscillation amplitude considered in our article the period reduction does not exceed 20\%. We conclude that the mass accretion can damp the motion of the threads rapidly. Thus, this mechanism can explain the observed strong damping of large-amplitude longitudinal oscillations. In addition, the damping time can be used to determine the mass accretion rate and indirectly the coronal heating.

}

\titlerunning{Damping by mass accretion}
\authorrunning{M. S. Ruderman \& M. Luna}

\maketitle

\section{Introduction}\label{sec:intro}

The first observation of a large-amplitude longitudinal oscillation (LALO) was first reported by \cite{jing2003}. In this oscillation a large portion of the cool prominence mass moved parallel to the filament axis with a total displacement of $140 \mathrm{\, Mm}$ and the velocity amplitude of $92 \mathrm{\, km\,s^{-1}}$\/, which is clearly in the range of large-amplitude motions according to the classification by \cite{oliver2002}. It seems that in this motion  the thread moved parallel to itself and to the local magnetic field. The oscillation period was about 80~min.\ with the damping time about 210~min., that is 2.6 times the period. These numbers indicate that the oscillation damping was very strong. More events have been reported later by \cite{jing2006}, \cite{vrsnak2007}, \cite{zhang2012}, \cite{Li2012}, \cite{luna2014}, \cite{bi2014}, and \citet{shen2014}. The range of the velocity amplitude was between $20$ and $100 \mathrm{\, km\,s^{-1}}$, the period between $40$ and $160\mathrm{\, min.}$, and the damping time between $1$ and $3.8$ periods. Again these observations indicate a very strong damping. The damping mechanism should be very efficient in order to damp very energetic LALOs so rapidly.

The origin of the prominence mass is nowadays an open question, but it has been known for a long time that the mass must come from the chromosphere \citep{pikelner1971}. It is unclear how the chromospheric mass is deposited into the corona. At present, the evaporation-condensation model \citep{antiochos1991} is the most advanced in its ability to explain thermal properties, speed, and mass of prominences \citep[e.g.,][]{antiochos2000,karpen2001,karpen2005,karpen2008,xia2011,luna2012a,xia2014, keppens2014}. In this model the coronal heating localized at the footpoints of the prominence magnetic structure produces the chromospheric plasma evaporation. This evaporated hot plasma flows along the field lines and condenses due to the optically thin radiation of the corona in a dipped parts of the magnetic field lines to form a cool prominence. Once the prominence is formed the same mechanism produces a constant accretion of mass into the prominence thread.

The damping mechanism of LALOs is poorly understood. Several damping mechanisms have been suggested but not rigorously tested, for example, the energy leakage by the sound wave emission \citep{kleczek1969} or some form of dissipation \citep{tripathi2009,oliver2009}. On the basis of numerical simulation \citep{luna2012a,luna2012b,luna2016} a model of the LALOs demonstrating that the restoring force is the projected gravity on magnetic tubes where the threads oscillate was constructed. In this model the motion is strongly damped by the steady accretion of mass onto the threads by the evaporation-condensation process. It was found that the temporal dependence of velocity of the thread is described by a Bessel function rather than by a sinusoid. This indicates that the accretion of mass by the threads not only damps the motion but also produces observable changes in the temporal profile of the oscillation. 

In the model suggested by \cite{luna2012b} the prominence threads were considered as point-like particles (0 dimension) with the increasing mass moving along a rigid field line. In this article we improve the model of \cite{luna2012b} by considering a 1D thread model moving in a rigid magnetic tube in the presence of an accretion flow.
The paper is organized as follows. In the next section the model is presented and the equations governing the thread motion are derived. In Sect.~\ref{sec:straight} the prominence thread oscillations in a magnetic tube consisting of a arc of a circle with two straight parts attached at its ends are considered. In Sect.~\ref{sec:circular} the thread oscillations in a magnetic tube with the shape of an arc of a circle are studied. Section~\ref{sec:sum} contains the summary of the results and our conclusions. 

\section{Derivation of governing equations}
\label{sec:derivat}

In the equilibrium there is a magnetic tube of constant cross-section. In Cartesian coordinates $x,\,y,\,z$ with the $z$\/-axis vertical the tube axis is in the $xz$\/-plane. Its shape is determined by the equations
\begin{equation}
x = x(s), \quad z = z(s),
\label{eq:1}
\end{equation}
where $s$ is the arclength measured along the axis. \cite{ruderman2015} showed that a magnetic tube with a constant cross-section radius and an arbitrary shape of its axis can be embedded in a potential magnetic field. Hence the functions $x(s)$ and $z(s)$ can be chosen arbitrarily. The gravity acceleration is $\vec{g} = (0,0,-g)$. The magnetic tube length is $\ell$\/, meaning that $0 \leq s \leq \ell$\/. There is a dense plasma with the density $\rho_p$ {\em per unit length} between $s = p$ and $s = q$\/. The plasma density  {\em per unit length} in $s < p$ and $s > q$ is $\rho_e < \rho_p$\/. Below we assume that $\rho_p$ and $\rho_e$ are constant. The unit tangent vector to the tube axis is $\vec{l} = (x'(s),0,z'(s))$, where the prime indicates the derivative. The projection of the gravity acceleration on the tube axis is $\vec{l}\cdot\vec{g} = -gz'(s)$. The projection on the tube axis of the gravity force acting on the element of the dense prominence thread from $s$ to $s + \Delta s$ is $-g\rho_p z'(s)\,\Delta s$\/. Then the total projection of the gravity force acting on the thread is
\begin{equation}
f_g = -\int_p^q g\rho_p z'(s)\,ds =  g\rho_p[z(p) - z(q)].
\label{eq:2}
\end{equation}

We assume that there is continuous plasma evaporation at the tube footpoints that creates the plasma flows at the two sides of the thread. The flow speed is $v = \mbox{const}$ and the flows are directed toward the thread at both sides. The plasma flux at both sides is the same and equal to $\rho_e v$\/. The thread velocity is $u$\/, and we assume that the plasma flow speed is larger than the thread speed, $v > |u|$. 

Our main assumption is that the accreting material is instantaneously absorbed by the thread, and its density, temperature and velocity become the same as those of the thread material. Due to accretion the velocity of the left end of the thread is smaller than $u$\/, while the velocity of the right end of the thread is larger than $u$\/. The relative velocity of the rarefied plasma flow and the thread is $v - \dot{p}$ at the left thread end, and $-(v + \dot{q})$ at the right thread end, where the dot indicates the time derivative. Hence, the rate of thread length increase at the left end is $\rho_e(v - \dot{p})/\rho_p$\/, while at the right end it is $\rho_e(v + \dot{q})/\rho_p$\/. Then it follows that
\begin{equation}
\dot{p} = u - \frac{\rho_e}{\rho_p}(v - \dot{p}), \quad 
\dot{q} = u + \frac{\rho_e}{\rho_p}(v + \dot{q}).
\label{eq:3}
\end{equation}
As a result, we obtain
\begin{equation}
\dot{p} = \frac{\rho_p u - \rho_e v}{\rho_p - \rho_e}, \quad
\dot{q} = \frac{\rho_p u + \rho_e v}{\rho_p - \rho_e}.
\label{eq:4}
\end{equation}
The mass of the thread is $M(t) = \rho_p(q - p)$. Differentiating this relation and using Eq.~(\ref{eq:4}) we obtain
\begin{equation}
\dot{M} = \rho_p(\dot{q} - \dot{p}) = \frac{2\rho_p\rho_e v}{\rho_p - \rho_e}. 
\label{eq:5}
\end{equation}
Integrating this equation yields
\begin{equation} 
M = \rho_p(q_0 - p_0) + \frac{2\rho_p\rho_e vt}{\rho_p - \rho_e} , 
\label{eq:6}
\end{equation}
where $p = p_0$ and $q = q_0$ at $t = 0$.

Consider a system consisting of the thread and two volumes of rarefied plasma attached to the thread at the left and the right. The length of the left volume is $(v - \dot{p})\Delta t$\/, while the length of the right volume is $(v + \dot{q})\Delta t$\/. The linear momentum of this system is
\begin{eqnarray}
M(t)u(t) + \rho_e v(v - \dot{p})\Delta t - \rho_e v(v + \dot{q})\Delta t \nonumber\\
= M(t)u(t) - \frac{2\rho_p\rho_e uv\,\Delta t}{\rho_p - \rho_e} .
\label{eq:6a}
\end{eqnarray}
After time $\Delta t$ the plasma in both volumes is absorbed by the thread. The thread mass and velocity become $M(t+\Delta t)$ and $u(t+\Delta t)$. The change in the linear momentum is equal to the impulse of force $f_g\,\Delta t$\/,
\begin{eqnarray}
M(t+\Delta t) u(t+\Delta t) &-& M(t)u(t) \nonumber\\
+\, \frac{2\rho_p\rho_e uv\,\Delta t}
{\rho_p - \rho_e} &=& g\rho_p[z(p) - z(q)]\Delta t .
\label{eq:6b}
\end{eqnarray}
Dividing this relation by $\Delta t$ and taking $\Delta t \to 0$ we obtain
\begin{equation} 
\frac{d(Mu)}{dt} = g\rho_p[z(p) - z(q)] - \frac{2\rho_p\rho_e u \, v}{\rho_p - \rho_e} .
\label{eq:7}
\end{equation}
Using Eq.~(\ref{eq:6}) we transform this equation to
\begin{equation} 
\left(q_0 - p_0 + \frac{2\rho_e vt}{\rho_p - \rho_e}\right)\dot{u} = 
   g[z(p) - z(q)] - \frac{4\rho_e uv}{\rho_p - \rho_e} .
\label{eq:8}
\end{equation}
Equations~(\ref{eq:4}) and (\ref{eq:8}) constitute the system of equations for $u$\/, $p$ and $q$\/.

{ We consider the model presented in this section as one of the first steps in understanding the damping mechanisms of LALOs, and we are well aware of its limitations. In this model we neglect many physical effect that, probably, exist in the reality. We consider the plasma motion in LALOs as one-dimensional while, in the reality, it is three-dimensional. The account of variation of the plasma parameters across the magnetic tube can result, for example, in the distortion of the boundary between the hot and cold plasmas which can complicate the process of hot plasma accretion. Probably, the most vulnerable assumption of our model is that the hot accreting plasma is instantaneously accommodated by the cold dense thread. Obviously the real process of hot plasma accretion is much more complex. The collision of flows of the hot and cold plasmas can cause formation of a very complex interaction region involving shocks. The relaxation of this region would, probably, involve heat conduction, ionisation and recombination. How much all these complicated processes will affect the damping of LALOs is an open question. It should be addresses in the future by considering more sophisticated models.}      

\section{Prominence oscillations in a magnetic tube with two straight parts}
\label{sec:straight}

\begin{figure}[ht]
\begin{center}
\includegraphics[width=0.4\textwidth]{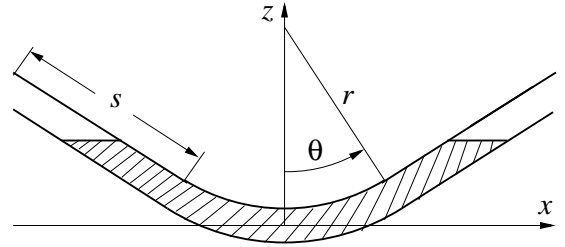} \vspace*{-3mm}
\end{center}
\caption{Sketch of the equilibrium. The magnetic tube axis consists of an arc of a circle of radius $r$ with the two straight lines attached. The dense prominence thread occupies the shaded area. { Recall that we consider a one-dimensional problem and neglect the variation of all quantities across the magnetic tube.}}  
\label{fig1} 
\end{figure}

\begin{figure}
\includegraphics[width=0.45\textwidth]{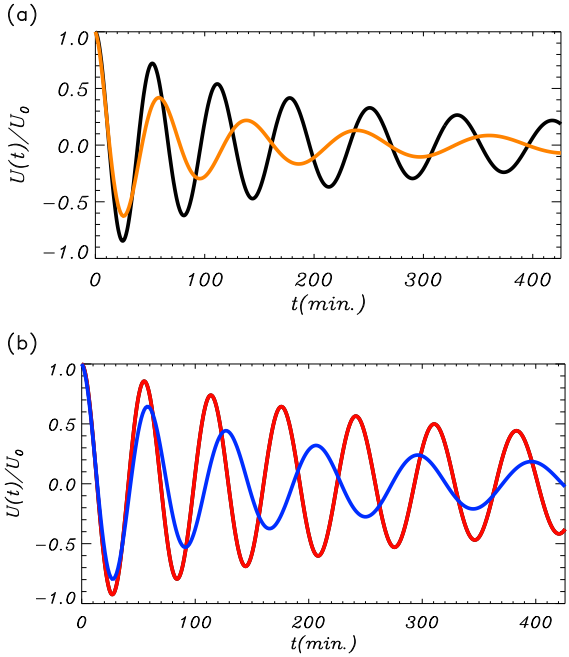}
\caption{Plot of the temporal evolution of the velocity of the thread given by Eq.~(\ref{eq:31}) normalized to the initial velocity $U_0$ as function of time, $t$. We have assumed a typical situation of $\zeta=100$, $r=60$ Mm, with $g = 274 ~\mathrm{m\, s^{-2}}$. For clarity, we have split the different cases studied in two panels. In (a) a thread with $l_p = 2.1$ Mm and $\theta=1^\circ$ is considered with $v = 10~\mathrm{km\, s^{-1}}$ (black curve) and $v = 30~\mathrm{km\, s^{-1}}$ (orange curve). In (b) a thread with $l_p = 5$ Mm and $\theta=3^\circ$ is considered with $v = 10~\mathrm{km\, s^{-1}}$ (red curve) and $v = 30~\mathrm{km\, s^{-1}}$ (blue curve).}
\label{fig:fig2} 
\end{figure}

To obtain the solution to the system of Eqs.~(\ref{eq:4}) and (\ref{eq:8}) we need to specify the magnetic tube shape. To make the problem as simple as possible we assume that the tube axis is composed of an arc of a circle with two straight lines attached to its ends such that the whole curve is smooth (see Fig.~\ref{fig1}). 
Hence, we take
\begin{equation} 
x(s) = \left\{\begin{array}{l}
(s + r\theta - \ell/2)\cos\theta - r\sin\theta , \\
\hspace*{30mm}   0 \leq s < \ell/2 - r\theta, \vspace*{2mm} \\
\displaystyle \hspace*{5mm} r\sin\frac{2s - \ell}{2r}, \qquad |s - \ell/2| \leq r\theta, \vspace*{2mm}\\
(s - r\theta - \ell/2)\cos\theta+ r\sin\theta , \\
\hspace*{30mm}   \ell/2 + r\theta < s \leq \ell, \end{array}\right.
\label{eq:13}
\end{equation}
\begin{equation} 
z(s) = \left\{\begin{array}{l}
(\ell/2 - s - r\theta)\sin\theta + r(1 - \cos\theta), \\ 
\hspace*{30mm}  0 \leq s < \ell/2 - r\theta, \vspace*{2mm} \\
\displaystyle r\left(1 - \cos\frac{2s - \ell}{2r}\right), \quad 
   |s - \ell/2| \leq r\theta, \vspace*{2mm} \\
(s - r\theta -\ell/2)\sin\theta + r\left(1 - \cos\theta\right), \\ 
\hspace*{30mm}  \ell/2 + r\theta < s \leq \ell, \end{array}\right.
\label{eq:14}
\end{equation}
where $\theta$ is the angle forming half of the arc with respect to the centre of curvature as plotted in Fig.~\ref{fig1}. We introduce the dimensionless variables
\begin{equation} 
\begin{array}{l} \displaystyle
\zeta = \frac{\rho_p}{\rho_e}, \quad P = \frac{gp}{v^2}, \quad 
Q = \frac{gq}{v^2}, \vspace*{1.5mm}\\ 
\displaystyle U = \frac uv, \quad T = \frac{gt}v, \quad Z = \frac{gz}{v^2}.
\end{array}
\label{eq:9}
\end{equation}
Then we rewrite the system of Eqs.~(\ref{eq:4}) and (\ref{eq:8}) in the dimensionless form as
\begin{equation} 
\left(Q_0 - P_0 + \frac{2T}{\zeta - 1}\right)\frac{dU}{dT} = 
   Z(P) - Z(Q) - \frac{4U}{\zeta - 1} ,
\label{eq:10}
\end{equation}
\begin{equation}
\frac{dP}{dT} = \frac{\zeta U - 1}{\zeta - 1}, \quad 
\frac{dQ}{dT} = \frac{\zeta U + 1}{\zeta - 1}.
\label{eq:11}
\end{equation}
Now we assume that $p < \ell/2 - r\theta$ and $q > \ell/2 + r \, \theta$ at any time, that is the ends of the dense thread are always on the straight parts of the magnetic tube. As a result, we obtain
\begin{equation}\begin{array}{cc}
Z(P) = \displaystyle \left(\frac L2 - P - R \, \theta\right)\sin\theta 
  + R\left(1 - \cos\theta\right), \vspace*{2mm} \\
Z(Q) = \displaystyle \left(Q - R \, \theta - \frac L2\right)\sin\theta 
  + R\left(1 - \cos\theta\right), \end{array}
\label{eq:15}
\end{equation}
where $L = g\ell/v^2$ and $R = gr/v^2$\/. It follows from this equation that
\begin{equation} 
Z(P) - Z(Q) = (L - P - Q)\sin\theta .
\label{eq:16}
\end{equation}
Substituting this result in Eq.~(\ref{eq:10}), differentiating the obtained equation, and using Eq.~(\ref{eq:11}), we obtain the equation for $U$\/, 
\begin{eqnarray} 
\frac d{dT}\left([(Q_0 - P_0)(\zeta - 1) + 2T]\frac{dU}{dT} + 4U\right) 
   = -2\zeta U\sin\theta . \nonumber\\ 
\label{eq:17}
\end{eqnarray}
Introducing the new variable $\sigma =  (Q_0 - P_0)(\zeta - 1) + 2T$ we rewrite this equation as
\begin{equation} 
\sigma\frac{d^2 U}{d\sigma^2} + 3\frac{dU}{d\sigma} + \kappa U = 0,  
\label{eq:18}
\end{equation}
where $\kappa = \frac12\zeta\sin\theta$\/. Below we assume that $Z(P_0) = Z(Q_0)$, which implies that $P_0 + Q_0 = L $. Then it follows from Eqs.~(\ref{eq:10}) and (\ref{eq:16}) that 
\begin{equation} 
(Q_0 - P_0)\frac{dU}{d\sigma} = - \frac{4U_0}{\zeta - 1} 
   \quad \mbox{at} \quad T = 0,
\label{eq:19}
\end{equation}
where $U_0 = U(0)$. As a result, we have the following initial conditions for $U$\/: 
\begin{equation}
U = U_0, \quad \frac{dU}{d\sigma} = - \frac{2U_0}{\sigma_0} 
\quad \mbox{at} \quad \sigma = \sigma_0,
\label{eq:20}
\end{equation}
where $\sigma_0 = (\zeta - 1)(Q_0 - P_0)$. The variable substitution
\begin{equation}
\xi = 2\sqrt{\kappa\sigma}, \quad U = \sigma^{-1} W,
\label{eq:21}
\end{equation}
reduces Eq.~(\ref{eq:20}) to the Bessel equation
\begin{equation} 
\frac{d^2 W}{d\xi^2} + \frac1\xi\frac{dW}{d\xi} + \left(1 - \frac4{\xi^2}\right)W = 0.
\label{eq:22}
\end{equation}
The general solution to this equation is
\begin{equation} 
W(\xi) = C_1 J_2(\xi) + C_2 Y_2(\xi),
\label{eq:23}
\end{equation}
where $J_2$ and $Y_2$ are the Bessel functions of the first and second kind, and $C_1$ and $C_2$ are arbitrary constants. Returning to the original variables we obtain
\begin{equation} 
U(\sigma) = \frac1\sigma\big[C_1 J_2\big(2\sqrt{\kappa\sigma}\big) + 
   C_2 Y_2\big(2\sqrt{\kappa\sigma}\big)\big],
\label{eq:24}
\end{equation}
Substituting Eq.~(\ref{eq:24}) in Eq.~(\ref{eq:20}) yields
\begin{eqnarray}
C_1 J_2\big(2\sqrt{\kappa\sigma_0}\big) + C_2 Y_2\big(2\sqrt{\kappa\sigma_0}\big) &=& \sigma_0 U_0, \vspace*{2mm} \\
\displaystyle C_1 J'_2\big(2\sqrt{\kappa\sigma_0}\big) + 
   C_2 Y'_2\big(2\sqrt{\kappa\sigma_0}\big) &=& -\sqrt{\frac{\sigma_0}\kappa}\,U_0, 
\label{eq:25}
\end{eqnarray}
where the prime indicates the derivative of Bessel function with respect to its argument. Using the identity \citep{abramowitz1972}
\begin{equation}
J_2(x) Y'_2(x) - J'_2(x) Y_2(x) = \frac2{\pi x},
\label{eq:26}
\end{equation}
we obtain from Eq.~(\ref{eq:25})
\begin{equation}
\begin{array}{l}
\displaystyle C_1 = \pi\sigma_0 U_0\big[\sqrt{\kappa\sigma_0}\, 
   Y'_2\big(2\sqrt{\kappa\sigma_0}\big) + Y_2\big(2\sqrt{\kappa\sigma_0}\big)\big], 
   \vspace*{2mm} \\
\displaystyle C_2 = -\pi\sigma_0 U_0\big[\sqrt{\kappa\sigma_0}\, 
   J'_2\big(2\sqrt{\kappa\sigma_0}\big) + J_2\big(2\sqrt{\kappa\sigma_0}\big)\big].
\end{array}
\label{eq:27}
\end{equation}
Typically $\zeta \sim 100$, while the initial length of the dense thread is of the order of a few Mm, $l_p = q_0 - p_0 \gtrsim 2$~Mm. If we take $v \lesssim 50$~km/s and $g  = 274$~m/s$^2$\/, then $Q_0 - P_0 \gtrsim 0.2$ and $\sigma_0 \gtrsim 20$. In addition, the radius of curvature could be estimated as of the order of $r = 60$~Mm \citep[see,][]{luna2014}. We are considering threads that are larger than the arched part of the tube, then $ l_p \ge 2 r \theta$\/.  For $l_p \gtrsim 2$~Mm this inequality can be satisfied if we take $\theta = 1^\circ$\/. Hence, below we assume that $\theta \gtrsim 1^\circ$\/. With these considerations $\kappa \gtrsim 0.8$. Taking into account that $\sigma \geq \sigma_0$ we arrive at the estimate $2\sqrt{\kappa\sigma} \gtrsim 8$. For such values of the argument we can use the asymptotic expressions for the Bessel functions \citep{abramowitz1972} namely
\begin{eqnarray}
&& J_m(x) \approx \sqrt{\frac2{\pi x}}\cos\left(x - \frac{\pi(2m+1)}4\right),  
\label{eq:28}\\
&& Y_m(x) \approx \sqrt{\frac2{\pi x}}\sin\left(x - \frac{\pi(2m+1)}4\right).
\label{eq:28a}
\end{eqnarray}
Then it follows from Eq.~(\ref{eq:27}) that
\begin{eqnarray}
&&C_1 \approx \frac{\sigma_0 U_0}\kappa\sqrt[4]{\pi^2\kappa\sigma_0}
   \cos\left(2\sqrt{\kappa\sigma_0} - \frac{5\pi}4\right), \quad
\label{eq:29}\\
&&C_2 \approx \frac{\sigma_0 U_0}\kappa\sqrt[4]{\pi^2\kappa\sigma_0}
   \sin\left(2\sqrt{\kappa\sigma_0} - \frac{5\pi}4\right).
\label{eq:29a}
\end{eqnarray}
Substituting Eqs.~(\ref{eq:29}) and (\ref{eq:29a}) in Eq.~(\ref{eq:24}) and using Eqs.~(\ref{eq:28}) and (\ref{eq:28a}) we transform it to the approximate form
\begin{equation} 
U(\sigma) = U_0\left(\frac{\sigma_0}\sigma\right)^{5/4}
   \cos\big(2\sqrt{\kappa\sigma} - 2\sqrt{\kappa\sigma_0}\big).
\label{eq:30}
\end{equation}
This equation can be rewritten as
\begin{eqnarray}
U(\sigma) &=& U_0\left(1 + \frac{T}{X}\right)^{-5/4} \nonumber\\
&\times& \cos\left(2\sqrt{2\kappa(X+T)} - 2\sqrt{2\kappa X}\right),
\label{eq:31}
\end{eqnarray}
where $X = \frac12(\zeta-1)(Q_0-P_0)$. In Fig.~\ref{fig:fig2} we have plotted four examples of the temporal evolution of the oscillating threads given by the previous equation.

The maximum displacement of the thread occurs when the argument of cosine is equal to $2\pi m$\/, $m = 0,1,\dots$ at times
\begin{equation} 
T = T_m \equiv \frac{\pi m}{2\kappa}\left(\pi m + 2 \sqrt{2\kappa X}\right).
\label{eq:32}
\end{equation}
Then the $n$-th oscillation period is given by
\begin{equation} 
\Pi_n = T_n - T_{n-1} = \frac\pi{2\kappa}\left[\pi(2n-1) + 2 \sqrt{2\kappa X}\right],
\label{eq:33}
\end{equation}
where $n = 1,2,\dots$. This indicates that the period of the oscillation depends of the cycle of the oscillation. In general this period, $\Pi_n$ increases with time. In Fig.~\ref{fig:fig2} the increase of the period for each oscillation is clear. The dimensional period $P_n$ is
\begin{eqnarray}
P_n &=& \frac vg\Pi_n = P_\mathrm{shift}+P_\mathrm{g} \nonumber\\
&=& \frac{v \pi^2 (2 n -1)}{g \zeta \sin \theta} +
    \pi \sqrt{\frac{2l_p(\zeta-1)}{g\zeta\sin \theta}} . 
\label{eq:period}
\end{eqnarray}

\noindent 
The second term in the expression for $P_n$ is associated with the gravity as a restoring force,
\begin{equation}\label{eq:pendulumperiod}
P_\mathrm{g} = \pi \sqrt{\frac{2l_p(\zeta-1)}{g\zeta\sin \theta}} .
\end{equation}
This term gives the oscillation period when $v = 0$. We can recover the oscillation period found by \citet{luna2012b} and \cite{luna2012c} if we assume that $\theta$ is small. Then $\sin \theta \approx \theta $ and
\begin{equation}
P_\mathrm{g} \approx \pi \sqrt{\frac{2l_p(\zeta-1)}{\theta \zeta g}} .
\end{equation}
In addition, \citet{luna2012b} and \cite{luna2012c} assumed that the thread filled the dipped part of the flux tube meaning that $r \theta = l_p/2$. Then, taking into account that, for typical prominences, the density contrast is very large meaning that $1-1/\zeta \sim 1$, we finally arrive at
\begin{equation}\label{eq:period-pendulum-approx}
P_\mathrm{g} \approx 2 \pi \sqrt{\frac{r}g} .
\end{equation}
Note that Eq.~(\ref{eq:period}) gives a more general expression. The first term in Eq.~(\ref{eq:period}), $P_\mathrm{shift}$, introduces the period shift. During each cycle of the oscillation the period increases by
\begin{equation}
\Delta P = \frac{2 v \pi^2}{g \zeta \sin \theta} .
\label{eq:shift}
\end{equation}
The period shift is related to the accretion rate onto the thread associated with the rarefied plasma. When the accretion rate increases so does the period shift. $P_\mathrm{shift} = 0$ when there is no accretion.

We define the damping time $T_d$ by the condition that the oscillation amplitude decreases $e$ times at $T = T_d$ with respect to the value at $T=0$ . Then, using Eq.~(\ref{eq:31}), we obtain
\begin{equation}
T_d = (e^{4/5} - 1)X = 2.12 X \, ,
\label{eq:34}
\end{equation}
and, in the dimensional variables,
\begin{equation}\label{eq:damping_time}
t_d = \frac vg T_d = 1.06(\zeta-1)\frac{l_p}{v} \, .
\end{equation}
This relation indicates that the oscillation damping time is inversely proportional to the accretion speed. The larger $v$ the smaller value of $t_d$ is, and, consequently, the stronger the damping is. Similarly, the larger the thread length $l_p$\/, the  weaker the damping is. The reason is that the term $(\zeta-1)l_p$ is essentially the mass of the thread at $t=0$. The damping occurs due to the increase of the thread mass and the decrease of its momentum. Hence, the larger the initial thread mass the more time is needed to damp its movement. 

In this section we have assumed that the cold thread is larger than the arched part of the tube, $l_p \geq 2 r \theta$. Typical prominence threads are equal or smaller than $10$ Mm but larger than a blob of $1$ Mm. In our previous studies we have found that $r$ should be of the order of tens Mm. In particular, in \citet{luna2014} we determined the radius of curvature of the dipped field lines of an observed filament as approximately equal to 60~Mm. Using these numbers we obtain that the angle $\theta \le 5^\circ$. The speed of the accretion flow, $v$\/, depends on the coronal heating at the flux-tube footpoints (see, e.g. \citealp{karpen2003}). Based on the simulations by \citet{luna2012a} and the results by \citet{karpen2005} we can estimate that the speed of the hot flows is of the order of $30 ~\mathrm{km \, s^{-1}}$. We also take a typical value of $\zeta = 100$ \citep[see][]{labrosse2010}. To plot Fig.~\ref{fig:fig2} we have considered four sets of parameters. Figure~\ref{fig:fig2}(a) corresponds to a thread of initial length $l_p = 2.1$~Mm and $\theta = 1^\circ$. The black line corresponds to the accretion velocity $v= 10 \mathrm{~km \, s^{-1}}$ and the orange line corresponds to $v= 30 \mathrm{~km \, s^{-1}}$. The difference in the accretion flow produces important changes in the damping time, $t_d$ (Eq. (\ref{eq:damping_time})), and in the period shift, $\Delta P$ (Eq. (\ref{eq:shift})), but not in the gravity period, $P_\mathrm{g}$ (Eq. (\ref{eq:pendulumperiod})). The damping and the period shift is stronger for the high accretion velocity (orange curve). For the case with $v=10 \mathrm{~km \, s^{-1}}$ (black line) $P_{\rm g}$ is 48.8 minutes, the shift is 3.4 minutes and the damping time  is 367 min., and for the case with $v=30 \mathrm{~km \, s^{-1}}$ (orange line) $P_{\rm g}$ is also 48.8 minutes, the shift is 10.3 minutes and the damping time is 122 minutes. Due to the shift the period defined as the time interval between two consecutive maxima changes for each oscillation. With Eq.~(\ref{eq:period}) we can compute the period of the $n$\/th cycle, as $P_n=3.4 (2 n-1) +48.8$ min. = 52.2, 55.7, 59.1, ... min. for the first set of parameters (black line). For the second set of parameters corresponding to the orange line the $n$\/th period is $P_n= 10.3 (2n -1) + 48.8$ min. = 59.1, 69.5, 79.8,\dots min. From this panel we clearly see the dependence of the damping time and the period shift on the accretion velocity. Larger values of the accretion velocity produces stronger damping, that is shorter damping times, and larger period shifts. A similar result can be seen in Fig.~\ref{fig:fig2}(b) for a larger thread of initial length $l_p = 5$~Mm. In this case the damping and the period shift is weaker than in the case with shorter threads (Fig.~\ref{fig:fig2}(a)). In fact, stronger damping, that is smaller damping time, involves larger period shifts. It is possible to combine Equations (\ref{eq:pendulumperiod}), (\ref{eq:shift}), and (\ref{eq:damping_time}) to obtain
\begin{equation}
t_d \, \Delta P = 1.06 \, P_\mathrm{g}^2 \, ,
\end{equation}
where we have assumed that the density contrast is large enough and taken $\zeta -1 \approx \zeta$\/. This relation reflects the fact that, for a given gravity period $P_{\rm g}$\/, a strong damping (small $t_d$) corresponds to a large period shift (large $\Delta P$) and vice versa. This behaviour is clear in both panels in Fig.~\ref{fig:fig2}. In Fig.~\ref{fig:fig2}(a) both cases have the same gravity period $P_\mathrm{g}$. The oscillation plotted in the orange curve has stronger damping and also larger period shift than in the case showed by the black curve. A similar effect can be seen in Fig.~\ref{fig:fig2}(b).

As we have already pointed out, it follows from Eq.~(\ref{eq:damping_time}) that the damping time is proportional to the initial length of the thread and inversely proportional to the speed of the accretion flow. This means that the damping is stronger for smaller initial threads and for stronger accretion flows. Using Eq.~(\ref{eq:6}) it is possible to relate the damping time with the mass of the thread at the initial time and the rate of mass accretion as
\begin{equation}\label{eq:damping-mass-1}
t_d = 2.12\, \frac{M (t=0)}{\dot{M}} \, ,
\end{equation}
This indicates that the damping is stronger in longitudinal oscillations produced in prominences with small thread mass.

\section{Prominence oscillations in a circular arched dip}
\label{sec:circular}

\begin{figure}[ht]
\begin{center}
\includegraphics[width=0.4\textwidth]{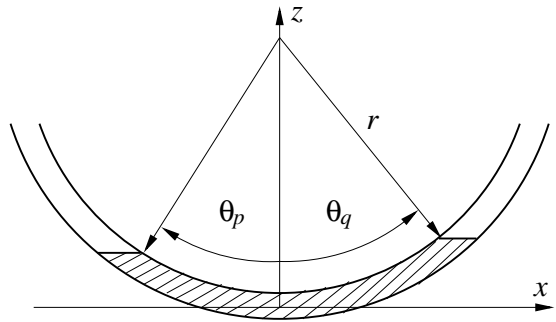} \vspace*{-3mm}
\end{center}
\caption{Sketch of the equilibrium. The magnetic tube axis is an arc of a circle of radius $r$\/. The prominence occupies the shaded area. { We again recall that we consider a one-dimensional problem and neglect the variation of all quantities across the magnetic tube.}}  
\label{fig3} 
\end{figure}

In this section we consider prominence oscillations in a magnetic tube that has the shape of an arc of a circle of radius $r$\/. The equilibrium state is shown in Fig.~\ref{fig3}. We introduce the angles $\theta_p$ and $\theta_q$ between the lines connecting the centre of the circular arc and the ends of the dense prominence thread. These angles are given by 
\begin{equation}
\theta_p = \frac{2p - \ell}{2r}, \quad \theta_q = \frac{2q - \ell}{2r} \, .
\label{eq:4.1}
\end{equation}
For this geometry we have
\begin{equation}
z(p) = r(1 - \cos\theta_p), \quad z(q) = r(1 - \cos\theta_q).
\label{eq:4.2}
\end{equation}
Introducing the dimensionless variables 
\begin{equation}
\tau = t\sqrt{\frac gr}, \quad \tilde{u} = \frac u{\sqrt{rg}}, 
\quad \tilde{v} = \frac v{\sqrt{rg}}, \quad \delta = \frac{l_p}r,
\label{eq:4.3}
\end{equation}
we can rewrite the system of Eqs.~(\ref{eq:4}) and (\ref{eq:8}) as
\begin{equation}
\frac{d\theta_p}{d\tau} = \frac{\zeta\tilde{u} -  \tilde{v}}{\zeta - 1}, \quad
\frac{d\theta_q}{d\tau} = \frac{\zeta\tilde{u} + \tilde{v}}{\zeta - 1},
\label{eq:4.4}
\end{equation}
\begin{equation} 
\left(\delta + \frac{2\tau\tilde{v}}{\zeta - 1}\right)\frac{d\tilde{u}}{d\tau} 
   = \cos\theta_q - \cos\theta_p - \frac{4\tilde{u}\tilde{v}}{\zeta - 1} .
\label{eq:4.5}
\end{equation}
Prominence threads have typical lengths of a few Mm and the radius of curvature of several tens of Mm. Then we can assume that the length of the thread is much smaller than the radius of the dip curvature, $l/r \ll 1$. This condition is equivalent to $\theta_q - \theta_p \ll 1$ meaning that we can use the approximate relation
\begin{eqnarray} 
\cos\theta_q - \cos\theta_p &=& -2\sin\frac{\theta_q + \theta_p}2
     \sin\frac{\theta_q - \theta_p}2 \nonumber\\
&\approx& -(\theta_q - \theta_p)\sin\phi ,
\label{eq:4.6}
\end{eqnarray}
where $\phi = (\theta_q + \theta_p)/2$. Since the typical value of $\zeta$ is 100, below we neglect 1 in comparison with $\zeta$\/. Then, using Eq.~(\ref{eq:4.6}), we
obtain from Eq.~(\ref{eq:4.4})
\begin{equation}
\tilde{u} = \frac{d\phi}{d\tau}, \quad 
\theta_q - \theta_p = \delta + \frac{2\tau\tilde{v}}\zeta .
\label{eq:4.7}
\end{equation}
With the aid of these results we reduce the Eq.~(\ref{eq:4.5}) to 
\begin{equation}
\frac{d^2\phi}{d\tau^2} + \sin\phi + \frac{4\tilde{v}}{\zeta\delta + 
   2\tau\tilde{v}}\frac{d\phi}{d\tau} = 0.
\label{eq:4.8}
\end{equation}
Below we assume that initially the dense thread is in equilibrium and then it is pushed and starts to oscillate. In accordance with this we impose the initial conditions
\begin{equation}
\phi = 0, \quad \frac{d\phi}{d\tau} = 2\chi_0
   \quad \mbox{at} \;\; \tau = 0,
\label{eq:4.8a}
\end{equation}
where $\chi_0$ is a constant related to the initial impulse given to the thread by some external trigger.

\subsection{Linear theory with strong damping}
\label{subsec:linear}

We first consider small-amplitude oscillations and assume that $\phi \ll 1$. Thus, we can use the approximate relation $\sin\phi \approx \phi$ and reduce Equation (\ref{eq:4.8}) to
\begin{equation}
\frac{d^2\phi}{d\tau^2} + \frac{4\tilde{v}}{\zeta\delta + 
   2\tau\tilde{v}}\frac{d\phi}{d\tau} + \phi = 0 \, .
\label{eq:L4.1}
\end{equation}
The variable substitution 
\begin{equation}
\xi = \tau + \frac{\zeta\delta}{2\tilde{v}}, \quad \phi = \xi^{-1}y \, ,
\label{eq:L4.2}
\end{equation}
reduces Equation (\ref{eq:L4.1}) to
\begin{equation}
\frac{d^2 y}{d\xi^2} + y = 0 .
\label{eq:L4.3}
\end{equation}
The general solution to this equation is a linear combination of $\sin\xi$ and $\cos\xi$\/. Then, returning to the original variables and using the initial conditions Eq.~(\ref{eq:4.8a}), we write the solution to Eq.~(\ref{eq:L4.1}) as
\begin{equation}
\phi (\tau) = \frac{2 \chi_0}{1 + 2\tilde{v}\tau/\zeta\delta}\sin\tau.
\label{eq:L4.4}
\end{equation}
This solution describes oscillations with constant period $\Pi = 2\pi$ in the dimensionless variables. Thus, in this model, the period is constant and, in the dimensional variables, it is given by
\begin{equation}\label{eq:pendulum-period-case2}
P=2 \, \pi \sqrt{\frac{r}{g}} \, ,
\end{equation}
which recovers the result by \cite{luna2012b}, \cite{luna2012c}, and \cite{luna2016}.
{ \cite{luna2016} numerically simulated the motion of perturbed localised cold plasma supported by a two-dimensional dipped magnetic field. They found that the back-reaction of the field on the plasma oscillation is very weak, validating the simpler assumption of rigid flux tubes in the present study. In particular, they obtained that the oscillation period was practically the same as that found by \cite{luna2012b}.} 

\begin{figure}[ht]
\begin{center}
\includegraphics[width=0.5\textwidth]{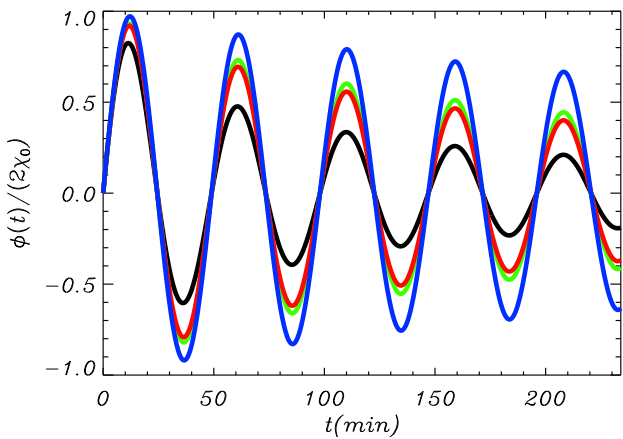}
\end{center}
\caption{Plot of the angular position $\phi (\tau)$ normalized to its amplitude $2\chi_0$ as a function of $t$ for various values of the parameters $v$ and $l_p$. We have taken $g = 274$~m\,s$^{-2}$ and typical values of $\zeta = 100$ and $r = 60$~Mm. In all cases the period is 49~min. The black curve corresponds to $v=30 \,\mathrm{km \, s^{-1}}$ and $l_p=2$~Mm, the green curve to $v=10 \,\mathrm{km \, s^{-1}}$ and $l_p=2$~Mm, the red curve to $v=40 \,\mathrm{km \, s^{-1}}$ and $l_p=5$~Mm, and the blue curve to $v=10 \,\mathrm{km \, s^{-1}}$ and $l_p=5$~Mm. The damping times for these combinations of parameters are $t_d=55.6,\,166.7,\,138.9,\,416.7$ min., respectively.} 
\label{fig:fig4} 
\end{figure}

The amplitude of the oscillations is given by the initial dimensionless velocity $2\chi_0 = u_0/\sqrt{r g}$\/, and the damping time depends of the factor $2\tilde{v}/(\zeta\delta)$. This solution implies that the damping rate changes with time similarly to what was found in the previous section. The damping is stronger at the initial stage of the oscillation close to $\tau=0$, then later it decreases for larger $\tau$\/. It is now convenient to introduce another dimensionless time $\Theta = vt/r = v\tau/\sqrt{gr}$\/. As in the previous section we define the dimensionless damping time $\Theta_d$ by the condition that the oscillation amplitude decreases $e$ times at $\Theta = \Theta_d$\/. Thus,
\begin{equation}
e^{-1}=\frac{\zeta\delta}{\zeta\delta + 2\tilde{v}\Theta_d} \, ,
\end{equation}
and we obtain
\begin{equation}\label{eq:dampingtimedimensionless_largeradius}
\Theta_d = (e-1) \frac{\zeta\delta}{2 \tilde{v}} \approx 
   0.86 \frac{\zeta\delta}{\tilde{v}} \, .
\end{equation}
In terms of dimensional variables we have
\begin{equation}\label{eq:dampingtimedimension_largeradius}
t_d=0.86 \, \frac{\zeta l_p}{v} \,,
\end{equation}
which implies that the damping is sufficiently strong for small threads and for the large accretion speed. Using Eq.~(\ref{eq:6}) it is possible to rewrite Eq.~(\ref{eq:dampingtimedimension_largeradius}) as
\begin{equation}\label{eq:damping-mass-2}
t_d=1.72 \, \frac{M (t=0)}{\dot{M}} \, ,
\end{equation}
which implies that the damping time not only depends on the mass accretion rate, but also on the initial mass of the thread. Equation~(\ref{eq:damping-mass-2}) is almost identical to Eq.~(\ref{eq:damping-mass-1}). They only differ by just a small difference in the constant at the front of the ratio of the initial mass and the mass accretion rate. Equation~(\ref{eq:dampingtimedimension_largeradius}) shows that strong damping is associated with large accretion rates and small initial thread masses. In Fig.~\ref{fig:fig4} we plot the temporal evolution of $\phi$ for several values of parameters $l_p$ and $v$\/. In all the cases the period is $P= 49$ minutes. We clearly see from this figure that the larger $l_p$ the weaker the damping is, that is the larger the damping time is. Similarly, the larger the accretion velocity the stronger the damping is, that is the smaller the damping time is.

\citet{luna2012b} found that the temporal evolution of the oscillation velocity was given by a Bessel function of order 1 and the damping is produced by the phase in the argument of this function. However, here we have found that the temporal evolution of the velocity is given by the sine divided by a linear function of its argument. The phase of the argument is one half of that found in the case studied by \citet{luna2012b}. The difference between the two models is in how the hot evaporated mass is deposited in the cool thread. \citet{luna2012b} assumed that the hot flows adapt to the motion of the thread and accretion at both sides of the thread is symmetric. Then, the momentum transferred to the thread by the hot evaporated flows is cancelled and the net momentum transfer is zero. In this case the damping is exclusively produced by the change of mass of the thread. In contrast, in the current work we assume that the hot evaporated flows are not affected by the motion of the thread. In this case there is a net transfer of the hot plasma flow momentum to the thread. This is given by the last term on the right hand side of Eq.~(\ref{eq:6a}). If we drop this term and solve the differential Eq.~(\ref{eq:8}), then we recover the temporal evolution found by \citet{luna2012b}. In a more realistic scenario that also includes the thermodynamic processes, the process of the momentum deposition is, probably, something in between these two extreme scenarios.

\subsection{Nonlinear weakly damped oscillations}
\label{subsec:linear}

When there is no accretion ($\tilde{v} = 0$) Eq.~(\ref{eq:4.8})  reduces to the equation of nonlinear pendulum. Its small-amplitude oscillation is described by $\phi(\tau) = \phi_0\sin\tau$, where $\phi_0$ is the constant oscillation amplitude. In that case the characteristic time of the variation of function $\phi(\tau)$ is 1. The characteristic time remains the same for nonlinear oscillations when the oscillation amplitude is smaller than or of the order of $\pi/2$\/. The damping of oscillations due to accretion can be considered as slow if the dimensionless damping time is much larger than 1. Since 1 is approximately equal to one sixth of the oscillation period, which is $2\pi$\/, this implies that the damping can be considered as slow if it is larger than or of the order of the oscillation period. This observation inspires us to search for a solution to the Eq.~(\ref{eq:4.8}) describing slowly damped nonlinear oscillations. To do this we introduce the ``slow'' time $\tau_1 = \epsilon\tau$\/, where $\epsilon \ll 1$ is of the order of the ratio of the characteristic oscillation time to the damping time. Then we consider $\phi$ as a function of two variables, $\tau$ and $\tau_1$\/. The damping is slow when the last term on the right-hand side of Eq.~(\ref{eq:4.8}) is small. In accordance with this we put $\tilde{v} = \epsilon\tilde{v}_1$\/. After that Eq.~(\ref{eq:4.8}) is transformed to
\begin{eqnarray}
\left(\frac{\partial^2\phi}{\partial\tau^2}\right. &+& \left. 
   2\epsilon\frac{\partial^2\phi}{\partial\tau\partial\tau_1} + 
   \epsilon^2\frac{\partial^2\phi}{\partial\tau_1^2}\right) + \sin\phi \nonumber\\
&+& \frac{4\epsilon\tilde{v}_1}{\zeta\delta + 2\tau_1\tilde{v}_1} 
   \left(\frac{\partial\phi}{\partial\tau} + 
   \epsilon\frac{\partial\phi}{\partial\tau}\right) = 0.
\label{eq:4.9}
\end{eqnarray}
Below we assume that $\phi$ is a periodic function of $\tau$ with the period $\Pi$ that will be determined later. Note that, in general, $\Pi$ can depend on $\tau_1$\/. We search for a solution to Eq.~(\ref{eq:4.9}) in the form of expansion
\begin{equation}
\phi = \phi_1 + \epsilon\phi_2 + \dots
\label{eq:4.10}
\end{equation}
Substituting this expansion in Eq.~(\ref{eq:4.9}) and collecting terms of the order of unity we obtain the equation of nonlinear pendulum
\begin{equation}
\frac{\partial^2\phi_1}{\partial\tau^2} + \sin\phi_1 = 0.
\label{eq:4.11}
\end{equation}
Using Eq.~(\ref{eq:4.8a}) we obtain the initial conditions for $\phi_1$\/,
\begin{equation}
\phi_1 = 0, \quad \frac{\partial\phi_1}{\partial\tau} = 2\chi_0 
   \quad \mbox{at} \;\; \tau = 0.
\label{eq:4.12}
\end{equation}
It is straightforward to obtain the first integral of Eq.~(\ref{eq:4.11}) satisfying the initial conditions Eq.~(\ref{eq:4.12}),
\begin{equation}
\left(\frac{\partial\phi_1}{\partial\tau}\right)^2 - 2\cos\phi_1 = 4\chi^2 - 2.
\label{eq:4.13}
\end{equation}
The quantity $\chi^2$ is proportional to the energy of the oscillation. When there is no damping the energy is conserved and $\chi = \chi_0$\/. However the energy decreases due to the damping, meaning that $\chi$ is a function of $\tau_1$\/. This function satisfies the initial condition $\chi = \chi_0$ at $\tau_1 = 0$.
 
The angle $\phi_1$ takes it maximum when $\partial\phi_1/\partial\tau = 0$. Then it follows from Eq.~(\ref{eq:4.13}) that the oscillation amplitude is 
\begin{equation}
A = \max\phi_1 = 2\arcsin\chi . 
\label{eq:4.13a}
\end{equation}
Below we assume that the oscillation amplitude does not exceed $\pi/2$\/. This condition implies that $\chi \leq \sqrt2/2$. We introduce the new dependent variable $\psi$ related to $\phi_1$ by
\begin{equation}
\sin\psi = \frac1\chi\sin\frac{\phi_1}2 , \quad -\frac\pi2 \leq \psi \leq \frac\pi2.
\label{eq:4.14}
\end{equation}
It follows from Eq.~(\ref{eq:4.13}) that the absolute value of the right-hand side of this equation does not exceed 1, so it always can be solved with respect to $\psi$\/. Now Eq.~(\ref{eq:4.13}) reduces to 
\begin{equation}
\left(\frac{\partial\psi}{\partial\tau}\right)^2 = 1 - \chi^2\sin^2\psi .
\label{eq:4.15}
\end{equation}
It follows from this equation that  
\begin{equation}
\tau = \int_0^\psi\frac{d\psi'}{\sqrt{1 - \chi^2\sin^2\psi'}},
\label{eq:4.16}
\end{equation}
where we have imposed the condition that $\psi$ is an increasing function of $\tau$\/, which corresponds to the first quarter of the first oscillation period. Then using the relation (\citealp{Korn1961}) ${\rm sn}(\tau;\chi) = \sin\psi$\/, where ${\rm sn}(\tau;\chi)$ is the elliptic sine, and Eq.~(\ref{eq:4.14}), we eventually obtain
\begin{equation}
\phi_1 = 2\arcsin(\chi\,{\rm sn}(\tau;\chi)).
\label{eq:4.17}
\end{equation}
This equation is valid for any $\tau \geq 0$. The oscillation period is four times the time needed for $\phi_1$ to vary from 0 to $A$\/. Since $\psi = 0$ when $\phi_1 = 0$ and $\psi = \pi/2$ when $\phi_1 = A$\/, it follows that the oscillation period is $\Pi = 4K(\chi)$, where $K(\chi)$ is the complete elliptic integral of the first kind given by (\citealp{Korn1961})
\begin{equation}
K(\chi) = \int_0^{\pi/2}\frac{d\psi}{\sqrt{1 - \chi^2\sin^2\psi}}.
\label{eq:4.18}
\end{equation}

To account for the effect of accretion we go to the next order approximation. Recall that now $\chi$ is a function of $\tau_1$\/. Collecting the terms of the order of $\epsilon$ in Eq.~(\ref{eq:4.9}) yields
\begin{equation}
\frac{\partial^2\phi_2}{\partial\tau^2} + \phi_2\cos\phi_1 = 
   -2\frac{\partial^2\phi_1}{\partial\tau\partial\tau_1} -
    \frac{4\tilde{v}_1}{\zeta\delta + 2\tau_1\tilde{v}_1} 
    \frac{\partial\phi_1}{\partial\tau} .
\label{eq:4.19}
\end{equation}
Since $\phi$ is a periodic function of $\tau$ with the period $\Pi$\/, the same is true for $\phi_2$\/. We multiply Eq.~(\ref{eq:4.19}) by $\partial\phi_1/\partial\tau$ and integrate with respect to $\tau$ from 0 to $\Pi$\/. Then, using Eq.~(\ref{eq:4.11}) and the integration by parts, we obtain on the left-hand side
\begin{eqnarray}
&&\int_0^\Pi\left(\frac{\partial^2\phi_2}{\partial\tau^2} + \phi_2\cos\phi_1\right)
   \frac{\partial\phi_1}{\partial\tau}d\tau \nonumber\\
&& \quad = \int_0^\Pi\phi_2\frac\partial{\partial\tau} \left(\frac{\partial^2\phi_1}
  {\partial\tau^2} + \sin\phi_1\right)d\tau = 0.
\label{eq:4.20}
\end{eqnarray}
This implies that the right-hand side is also zero, which gives the equation
\begin{eqnarray} 
\frac d{d\tau_1}\int_0^\Pi\left(\frac{\partial\psi_1}{\partial\tau}\right)^2 d\tau + 
   \frac{4\zeta\tilde{v}_1}{\zeta\delta + 2\tau_1\tilde{v}_1} 
   \int_0^\Pi\left(\frac{\partial\psi_1}{\partial\tau}\right)^2 d\tau = 0. \nonumber\\
\label{eq:4.21}
\end{eqnarray}
In this equation we use the ordinary derivative because the integral in this equation only depends on $\tau_1$\/. Using Eq.~(\ref{eq:4.13}) yields
\begin{equation}
\int_0^\Pi\left(\frac{\partial\psi_1}{\partial\tau}\right)^2 d\tau = 
   4\int_0^\Pi\left(\chi^2 - \sin^2\frac{\phi_1}2\right)d\tau .
\label{eq:4.22}
\end{equation}
Then, with the aid of Eqs.~(\ref{eq:4.14}) and (\ref{eq:4.16}) we obtain 
\begin{equation}
\int_0^\Pi\left(\frac{\partial\psi_1}{\partial\tau}\right)^2 d\tau = 
   4\chi^2\int_0^\Pi\cos^2\psi\,d\tau = 16\Upsilon(\chi),
\label{eq:4.23}
\end{equation}
where
\begin{eqnarray}
\Upsilon(\chi) &=& \chi^2\int_0^{\pi/2}\frac{\cos^2\psi\,d\psi}
   {\sqrt{1 - \chi^2\sin^2\psi}} \nonumber\\
&=& E(\chi) - (1 - \chi^2)K(\chi), 
\label{eq:4.24}
\end{eqnarray}
and the complete elliptic integral of the second kind $E(\chi)$ is given by (\citealp{Korn1961})
\begin{equation}
E(\chi) = \int_0^{\pi/2}\frac{d\psi}{\sqrt{1 - \chi^2\sin^2\psi}}.
\label{eq:4.25}
\end{equation}
Using Eq.~(\ref{eq:4.23}) we transform Eq.~(\ref{eq:4.21}) to
\begin{equation}
\frac{d\Upsilon(\chi)}{d\tau_1} + \frac{4\tilde{v}_1\Upsilon(\chi)} 
   {\zeta\delta + 2\tau_1\tilde{v}_1} = 0.
\label{eq:4.26}
\end{equation}
It follows from this equation that 
\begin{equation}
\Upsilon(\chi) = \frac{\zeta^2\delta^2\Upsilon(\chi_0)}
   {(\zeta\delta + 2\Theta)^2}.
\label{eq:4.27}
\end{equation}
Recall that $\Theta = vt/r = \tau_1\tilde{v}_1 = \tau\tilde{v}$\/, and $\chi_0$ is the value of $\chi$ at the initial time ($\tau_1 = 0$). It follows from Eq.~(\ref{eq:4.24}) that $\Upsilon(\chi)$ is a monotonically increasing function. Then it follows from Eq.~(\ref{eq:4.27}) that $\chi$ decreases with time. Using the expression for the oscillation amplitude $A$ in terms of $\chi$ we conclude that $A$ also decreases with time, which is an expected result. Again we define the dimensionless damping time $\Theta_d$ as the time when the oscillation amplitude becomes $e$ times smaller than the initial amplitude $A_0$\/. Using Eqs.~(\ref{eq:4.13a}) and (\ref{eq:4.27}) we obtain
\begin{equation}
\Theta_d = \frac{\zeta\delta}2\left(\sqrt{\frac{\Upsilon(\sin(A_0/2))}  
   {\Upsilon(\sin(A_0/2e))}} - 1\right)  . 
\label{eq:4.28}
\end{equation}
Rewriting this expression in the dimensional variables gives the expression for the dimensional damping time $t_d$\/,
\begin{equation}
t_d = \frac{\zeta l_p}{2v}
   \left(\sqrt{\frac{\Upsilon(\sin(A_0/2))}
   {\Upsilon(\sin(A_0/2e))}} - 1\right)  . 
\label{eq:4.29}
\end{equation}

The theory becomes especially simple in the linear approximation that we obtain assuming that $\chi \ll 1$. Then $A = 2\chi$\/, ${\rm sn}(\tau;\chi) = \sin\tau$\/, 
and $K(\chi) = E(\chi) = \pi/2$. Using these relations and Eq.~(\ref{eq:4.24}) we obtain that $\Pi = 2\pi$ and $\Upsilon = \pi\chi^2/2$. Now we obtain from Eqs.~(\ref{eq:4.27}) and (\ref{eq:4.29}) that
\begin{equation}
\chi = \frac{\zeta\delta\chi_0}{\zeta\delta + 2vt/r}, \quad
t_d = \frac{\zeta l_p(e-1)}{2v}=0.86 \frac{\zeta l_p}{v}.
\label{eq:4.30}
\end{equation}
Finally, it follows from Eq.~(\ref{eq:4.27}) 
\begin{equation}
\phi_1 = 2\chi\sin\tau = 
\frac{\zeta l_p A_0}{\zeta l_p + 2vt}\sin\big(t\sqrt{g/r}\big),
\label{eq:4.31}
\end{equation}
where $A_0 = 2\chi_0$\/. We see that the expression for $t_d$ coincides with that given by Eq.~(\ref{eq:dampingtimedimensionless_largeradius}). It is also straightforward to verify that Eq.~(\ref{eq:4.31}) coincides with Eq.~(\ref{eq:L4.4}). Hence, we recovered the results obtained in Subsection~\ref{subsec:linear}. 

\begin{figure}[ht]
\begin{center}
\includegraphics[width=0.45\textwidth]{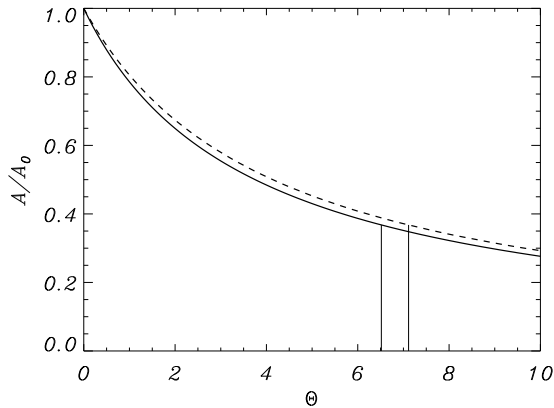} \vspace*{-5mm}
\end{center}
\caption{Dependence of the oscillation amplitude $A$ on the dimensionless time $\Theta = vt/r$\/. The solid and dashed curves correspond to the initial amplitude $A_0 = \pi/2$ and $A_0 = \pi/8$\/, respectively. The vertical lines indicate the damping time $\Theta_d$\/.}  
\label{fig:fig5} 
\end{figure}

In Fig.~\ref{fig:fig5} the dependence of $A$ on $\Theta$ for $\zeta = 100$, $\delta = 1/12$, and two values of the initial amplitude $A_0 = 2\arcsin\chi_0$, $A_0 = \pi/8$ and $A_0 = \pi/2$, are shown. We did not show the curve obtained using the linear theory because it practically coincides with that corresponding to $A_0 = \pi/8$. We see that the nonlinearity only slightly reduces the damping time. For $\zeta = 100$ and $\delta = 1/12$ the linear theory gives $\Theta_d = 7.16$, while the nonlinear theory gives $\Theta_d = 7.11$ when $A_0 = \pi/8$ and $\Theta_d = 6.52$ when $A_0 = \pi/2$. Hence, even when $A_0 = \pi/2$ the nonlinearity reduces the damping time by less than 10\%.

When there is no damping the oscillation period is equal to $4K(\chi_0)$. However, due to damping $\chi$ decreases with time.

Consider the sequence $\{\tau_n\}$\/, $n = 0,1,\dots$, where $\phi = 0$ at $\tau = \tau_{2n}$\/, $\phi$ takes its local maximum when $\tau = \tau_{2n+1}$\/, and it takes its local minimum when $\tau = \tau_{2n+3}$\/. The $n$\/th oscillation cycle corresponds to the variation of $\tau$ from $\tau_{2n-2}$ to $\tau_{2n+2}$\/. The angle $\phi$ increases from 0 to its local maximum when $\tau$ varies from $\tau_{2n}$ to $\tau_{2n+1}$\/, than it decreases back to 0 when $\tau$ varies from $\tau_{2n+1}$ to $\tau_{2n+2}$\/, continues to decrease to reach its local minimum when $\tau$ varies from $\tau_{2n+2}$ to $\tau_{2n+3}$\/, and finally return to 0 when $\tau$ varies from $\tau_{2n+3}$ to $\tau_{2n+4}$\/. Hence, we split each oscillation period in four quarters. Since $\chi$ is a slowly varying function of $\tau$ we can neglect it variation in any quarter of period. Each quarter of period corresponds to the variation of $\psi$ by $\pi/2$\/. Then, using Eqs.~(\ref{eq:4.16}) and (\ref{eq:4.18}), we obtain the recurrence relation  
\begin{equation}
\tau_{n+1} - \tau_n = \int_0^{\pi/2}\frac{d\psi}
   {\sqrt{1 - \chi^2(\tau_n)\sin^2\psi}} = K(\chi(\tau_n)).
\label{eq:4.32}
\end{equation}
The $n$\/th oscillation period is given by
\begin{equation}
\Pi_n = \tau_{4n+4} - \tau_{4n} = \sum_{j=0}^3 K(\chi(\tau_{4n+j})).
\label{eq:4.33}
\end{equation}
The function $\chi(\tau)$ is defined by Eq.~(\ref{eq:4.27}).

Since $\chi$ is a monotonically decreasing function of $\tau$ and $K(\chi)$ is a monotonically decreasing function of $\chi$\/, it follows that $\{\Pi_n\}$ is a monotonically decreasing sequence. When $\tau \to \infty$, $\chi \to 0$\/, $K(\chi) \to \pi/2$\/, and $\Pi_n \to 2\pi$\/. The stronger the damping the faster $\chi(\tau)$ decreases and, consequently, the faster the sequence $\{\Pi_n\}$ decreases. The larger the initial amplitude $A_0$ the larger $\chi_0$ is and, consequently, the larger the difference between the initial period, $\Pi_1$\/, and the limiting period value $2\pi$ is. However this difference is not very big even for quite larger initial oscillation amplitude. When $A_0 = \pi/2$ we obtain $\Pi_1 < 4K(\chi_0) \approx 7.42$, meaning that, even for this large value of the oscillation amplitude, the difference between $\Pi_1$ and $2\pi$ is less than 20\%.

As an example, using Eqs.~(\ref{eq:4.27}), (\ref{eq:4.32}), and (\ref{eq:4.33}) we calculated oscillation periods $P_n = \Pi_n\sqrt{r/g}$ for $A_0 = \pi/2$\/,    $\zeta = 100$\/, $g = 274~\mathrm{m\,s^{-2}}$\/, $r = 60~\mathrm{Mm}$, $l_p = 2~\mathrm{Mm}$, and $v = 30~\mathrm{km\,s^{-1}}$\/. We obtained $P_1 = 53.9~\mathrm{min.}$, $P_2 = 50.4~\mathrm{min.}$, $P_3 = 49.7~\mathrm{min.}$, and $P_4 = 49.4~\mathrm{min.}$. $P_n \to 2\pi\sqrt{r/g} = 49~\mathrm{min.}$ as $n \to \infty$\/. Hence, in this particular example the period only decreases by 10\%.

\begin{figure}
\includegraphics[width=0.45\textwidth]{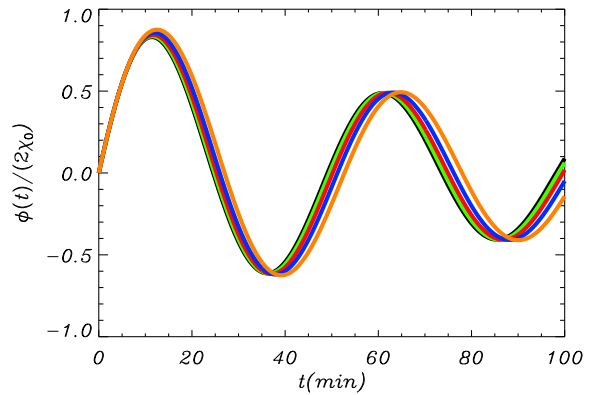}
\caption{Plot of the temporal evolution of the angle $\phi$ described by Eq. (\ref{eq:4.8}) normalized to the initial dimensionless velocity $2\chi_0$ as function of time, $t$\/. We have taken $g = 274 ~\mathrm{m\, s^{-2}}$ and typical values of $\zeta=100$, $r=60$~Mm, and $l_p = 2$~Mm. The black, green, red, blue, and orange correspond to the initial velocities of $u_0=2\chi_0\sqrt{rg} = 36, 72, 108, 144$, and $180 ~\mathrm{km \, s^{-1}}$\/, respectively. }
\label{fig:fig6} 
\end{figure}

In Figure \ref{fig:fig6} we have plotted the full numerical solutions for typical values of parameters. We clearly see the nonlinear effects but also we see that these effects are not significant. The orange curve corresponds to $A_0 = \pi/2$ that in dimensional variables corresponds to the initial velocity equal to $180~\mathrm{km\,s^{-1}}$. We see that it is only slightly different from the black curve corresponding to the initial velocity equal to $36~\mathrm{km\,s^{-1}}$.  

\section{Summary and conclusion}
\label{sec:sum}

In this article we have studied the damping of longitudinal oscillations of a prominence thread caused by the mass accretion of the evaporated chromospheric plasma. We have considered a thin curved magnetic tube of an arbitrary shape. The prominence thread is in the central part of the tube and it consists of a dense cold plasma. The parts of the tube at the two sides of the thread are filled with a hot rarefied plasma. The restoring force in the prominence oscillation is the gravity projected on the flux tube. We have assumed that there are flows of coronal rarefied plasma toward the thread. These flows are caused by the plasma evaporation at the magnetic tube footpoints. The coronal heating is localized at the chromosphere and at the bottom of the corona produces the evaporation. The hot evaporated plasma condenses in the already formed prominence thread by the thermal non-equilibrium instability. Our main assumption is that the hot evaporated plasma is instantaneously accommodated by the thread when it arrives at the thread, and its temperature and density become equal to those of the thread. Then we derived the system of three ordinary differential equations describing the thread dynamics.

The equations describing the thread oscillation are valid for an arbitrary shape of the magnetic tube axis. The only restriction is that it is a planar curve in a vertical plane. Of course the oscillation properties depend on a particular shape of the magnetic tube. We considered two particular models. In the first one the magnetic tube axis is composed of an arc of a circle with two straight lines attached to its ends in such a way that the whole curve is smooth. A very important property of this model is that the equations describing the thread oscillations are linear for any oscillation amplitude under the restriction that the thread ends remain of the straight parts of the tube. We obtained the solution to the governing equations in terms of Bessel functions. We showed that, for typical parameters of solar prominences, the Bessel functions can be approximated by trigonometric functions. Then we obtained the analytical expressions for the  oscillation damping time and periods. We found that the damping time is inversely proportional to the accretion rate and proportional to the initial mass of the thread. The oscillation period depends strongly on the angle between the straight parts of the tube axis and the horizontal direction. The larger this angle, the smaller the period is. We also found that the period increases with time and, in each cycle, the time of the maximum thread displacement is shifted. We have found that the larger the damping the larger the period shift for a given oscillation period.

In the second model studied in this article the shape of the tube axis is an arc of a circle. We have considered the linear as well as the nonlinear regime. In the linear regime we assumed that the displacement of the thread is small in comparison with the radius of curvature of the dipped flux tube. We have found that the period is equal to the period of the pendulum oscillation and it does not change with time. The damping time is inversely proportional to the mass accretion rate and proportional to the initial mass of the thread. In the nonlinear regime, we assumed that the damping is slow meaning that the damping time is much larger that the characteristic oscillation time. It is important to note that the characteristic oscillation time is the oscillation period divided by $2\pi$\/. This implies that the damping can be considered as slow even when the damping time is of the order of the oscillation period. To study the thread oscillations we used the two-scale approach where the oscillations are described by the solution of the nonlinear pendulum problem with slowly varying amplitude. We showed that the nonlinearity only slightly reduces the damping time. Again the damping time is inversely proportional to the accretion speed and proportional to the initial mass. In this model the oscillation periods decrease with time. This behaviour is in contrast with that found in the first model. The larger the initial oscillation amplitude the larger the reduction in the oscillation periods is. However, even for the largest initial oscillation amplitude considered in our article this reduction does not exceed 20\%.      

We conclude that the mass accretion can damp the motion of the threads rapidly. Thus, this mechanism can explain the observed strong damping of large-amplitude longitudinal oscillations. In addition, the damping time can be used to determine the mass accretion rate and indirectly the coronal heating. More work needs to be done to increase the complexity of the model by including stratification of the plasma, the physical processes in condensation of the thermal instability, and consider 2D and 3D models of the magnetic geometry in order to understand the interaction of the plasma with the magnetic field. In addition, the damping by radiative losses should be considered in a full model. \citet{Zhang2013a} found that effect of the radiative losses can be significant in these oscillation. Recently, \citet{ballester2016} have found that a temporal variation of the background temperature in combination with radiative losses can produce period shifts and damping of the slow modes in a prominence. These improvements to the model will be a topic for future research.

\acknowledgement
This paper was inspired by two ISSI workshops, Bern, Switzerland, March and
November 2015. The authors acknowledge support from the International Space Science Institute (ISSI) to the Team 314 on ``Large-Amplitude Oscillation in prominences'' led by M. Luna. MR acknowledges the financial support from the Science and Technology Facilities Council (STFC). M. Luna acknowledges the support by the Spanish Ministry of Economy and Competitiveness through projects AYA2011-24808, AYA2010-18029 and AYA2014-55078-P. This work contributes to the deliverables identified in FP7 European Research Council grant agreement 277829, ``Magnetic Connectivity through the Solar Partially Ionized Atmosphere'' (PI: E. Khomenko).

\bibliographystyle{aa}
\bibliography{ruderman}
\end{document}